# Enabling remote quantum emission in 2D semiconductors via porous metallic networks

*Jose J. Fonseca[1]\*, Andrew L. Yeats[1], Brandon Blue[2], Maxim Zalalutdinov[1], Todd Brintlinger[1], Blake S. Simpkins[1], Daniel C. Ratchford[1], James C. Culbertson[1], Joel Q. Grim[1], Samuel G. Carter[1], Masa Ishigami[2], Rhonda M. Stroud[1], Cory Cress[1], Jeremy T. Robinson[1]\**

[1] Naval Research Laboratory, Washington, DC 20375, USA
[2] Department of Physics and Nanoscience Technology Center, University of Central Florida, Orlando, FL 32816

\*Corresponding authors

**Abstract**

The interaction between two-dimensional crystals (2DCs) and metals is ubiquitous in 2D material research. Here we report how 2DC overlayers influence the recrystallization of relatively thick metal films and the subsequent synergetic benefits this provides for coupling surface plasmon-polaritons (SPPs) to photon emission in 2D semiconductors. We show that annealing 2DC/Au films on $SiO_2$ results in a 'reverse epitaxial' process where initially nanocrystalline Au films become highly textured and in close crystallographic registry to the 2D crystal overlayer. With continued annealing, the metal underlayer dewets to form an oriented pore enabled network (OPEN) film in which the 2DC overlayer remains suspended above or coats the inside of the metal pores. This OPEN film geometry supports SPPs launched by either direct laser excitation or by light emitted from the TMD semiconductor itself, where energy in-coupling and out-coupling occurs at the metal pore sites such that dielectric spacers between the metal and 2DC layer are unnecessary. At low temperatures a high density of single-photon emitters (SPEs) is present across an OPEN-$WSe_2$ film, and we demonstrate non-local excitation of SPEs at a distance of 17 μm with minimal loss of photon purity. Our results suggest the OPEN film geometry is a versatile platform that could facilitate the use of layered materials in quantum optics systems.

**Introduction**

Integrated quantum optics platforms using solid-state quantum emitters have motivated the search for ideal on-chip single photon sources that can be interfaced with one another to form networks for quantum technologies such as computing and communication. 2D semiconductors may provide such sources, with facile integration of on-chip elements due to their intrinsic immunity to total internal reflection and the possibility of van der Waals epitaxy on a wide range of substrates. The initial finding of random single photon emitters (SPEs) in WSe$_2$[1-5] quickly led to the deterministic placement of SPEs with narrow linewidths.[6-9] While the exact nature of the emitter sites remains unknown, the general consensus is that both defects[2-5,10] and strain[11,12] are sufficient to localize excitons with very narrow radiative line widths. When paired with metals, these emitters can be further engineered by, for example, a cascade energy transfer process[13] where excitons in a 2D semiconductor can excite surface plasmon-polaritons (SPPs) and vice versa.[14] This opens opportunities for coupling and/or energy routing between SPE sites in 2D semiconductors for potential applications such as on-chip quantum light sources.[15-17] In addition, metal nanostructures can lead to enhanced Purcell factors resulting in increased single-photon emission rates.[9,18] As such, it is natural to imagine interfacing 2D semiconductor SPEs with metals to advance quantum and optical processing technology.

Like most surfaces and interfaces, the 2DC/metal junction dictates the ultimate performance of an engineered solid-state system. For example, assembling 2DC/metal junctions by transfer stacking, versus direct metal deposition, reduces both chemical disorder and Fermi level pinning at the interface, to the point of approaching the Schottky–Mott limit for a van der Waals metal–semiconductor junction.[19] Alternatively, 2DC/metal interfaces with crystallographic registry can be formed through van der Waals epitaxy.[20] During physical vapor deposition, thin metal films tend to preferentially align on 2DCs,[21,22] which results in periodic (Moiré) structural and electronic variations at the interface due to the lattice mismatch. We demonstrate a platform that brings together concepts from each of these areas: aligned 2DC/metal interfaces, SPP-exciton coupling, and 2D quantum emitters.

In this work, we exploit dewetting of 2DC/metal films to assemble structures that can be used for energy propagation and energy transfer between a 2D semiconductor and a metal film. The relative surface energy of a thin metal film on a substrate determines the driving force for wetting or dewetting.[23] Despite being atomically thin, 2D materials significantly modify metal film dewetting due to their relatively high Young's Modulus, which suppresses surface fluctuations.[24] We show here that under appropriate annealing conditions, the metal underlayer becomes highly textured and in crystallographic alignment with the 2DC overlayer, analogous to a 'reverse epitaxy' process (or surface-templated epitaxy). As dewetting initiates, pores form in the metal film while the 2DC overlayer either remains suspended above or coats the inside of the pores. We term this 2DC/metal framework an orientated pore enabled network (OPEN) film. In this OPEN film geometry, energy in-coupling/out-coupling between SPPs in the metal and excitons in the 2D semiconductor occur at the metal pore sites. We highlight that the metal pore site serves as an air or vacuum dielectric spacer, which is analogous to typical sample geometries that use solid-state dielectric spacers between the metal and 2DC layer (e.g., h-BN,[14] $Al_2O_3$ [9,13]). Furthermore, the porous Au framework connects neighboring SPEs located at the pore sites, thereby enabling the 'networking' aspect of the OPEN film geometry.

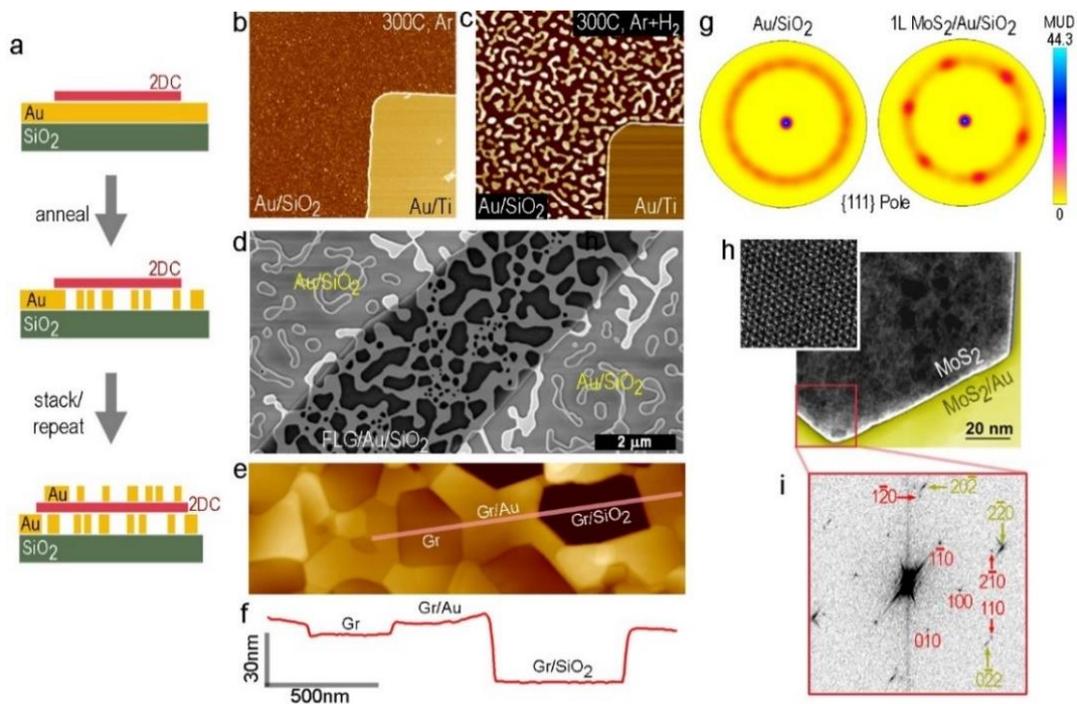

**Figure 1 | Fabrication and characterization of gold films (25nm) with 2DC overlayers.** (**a**) Schematic of sample fabrication and processing. (**b, c**) AFM images showing Au + patterned Au/Ti layers on $SiO_2$ after annealing in flowing (**b**) Ar and (**c**) Ar+$H_2$ at 300 °C (image: 14×14μm$^2$). (**d**) SEM image of an exfoliated few-layer graphene flake on Au/$SiO_2$ after annealing. (**e**) AFM image of a CVD graphene monolayer on de-wetted Au pores. (**f**) Cross-sectional height profile from the line drawn in **e** showing an example of a suspended graphene (Gr) membrane and a graphene membrane coating the inside of a pore. (**g**) EBSD {111} pole figures from annealed Au/ $SiO_2$ and

annealed monolayer-MoS$_2$/Au/SiO$_2$ (scale = multiple of uniform density (MUD)). (**h**) High-resolution STEM imaging and (**i**) corresponding FFT diffractogram showing the relative lattice orientation of a MoS$_2$ membrane (red indices) anchored to a de-wetted Au support (gold indices).

## Results

**2DC/metal architectures.** The basic structure for our 2DC/metal systems consist of a 2DC layer transferred onto a thin metal film without the use of a metal adhesion layer (Fig. 1a and Methods section). The simplicity of the process allows for the use of various 2DC layers or metal layers in a wide variety of combinations. In this work we focus on Au thin films on SiO$_2$ and a variety of 2DC material overlayers, but the process appears to be general, working with other metal films (see SI Fig. S1). The Au film properties, substrate and annealing conditions were selected to promote metal dewetting at low temperatures even in the absence of the 2DC layer. We find the addition of hydrogen gas to the annealing ambient reduces the onset temperature of dewetting by at least 100°C as compared to Ar alone (Fig. 1b,c), presumably through the reduction of suboxides and/or local carbonaceous pinning sites.

The dramatic effect a 2DC overlayer has on Au dewetting is highlighted in Figure 1d. In this example, a narrow exfoliated strip of few-layer graphene was deposited on a 25 nm thick Au film and the sample was annealed at 350 °C for 30 min. As thermodynamic modeling suggests,[24] Au film dewetting is impeded beneath the graphene layer as compared with the Au-only region. Unlike the full prevention of Au dewetting reported by Cao et al.,[24] we find reduced dewetting under the graphene resulting in metal pores (dark regions, Fig. 1d) that range in diameter from approximately 100-1000 nm depending on annealing conditions. Atomic force microscopy (AFM) of this and other samples (e.g., Fig. 1e) confirm that the 2DC layers can remain suspended above the metal pores with close to 100 % yield (forming tens-of-thousands of nanomechanical membranes in one step), despite the strain involved as the metal underlayer moves. Gold diffusion during dewetting leads to local variations in film height of up to about ±5 nm (dependent on the extent of pore coverage), which influences the resultant shape of and tension within the suspended 2DC membranes (Fig 1e,f). As shown in Fig. 1f, the suspended membranes clamp along the top interior of the pore wall, which is commonly observed in strained 2DC membranes.[25]

The 2DC overlayer not only impedes the metal dewetting process, it strongly affects the recrystallization of the Au film itself. As seen in the AFM image in Fig. 1e, the Au underlayer forms faceted pores and the Au surface has clearly resolvable step boundaries, suggesting a transformation from a nano- to micro-crystalline film. Using electron backscattered diffraction (EBSD; see Methods) we assess how the Au film recrystallizes with and without a 2DC overlayer. Figure 1g (and SI Fig. S2) shows a {111} pole plot from two annealed samples: $Au/SiO_2$ and monolayer-$MoS_2/Au/SiO_2$. We specifically choose exfoliated material here to have confidence of a single-crystal 2DC domain across the entire analyzed area (60x40 $\mu m^2$) and note that EBSD is insensitive to the 2DC monolayer. The pole plots show that the Au film becomes highly textured in the presence of $MoS_2$ after annealing (300 °C, 1 hour), with two {111} in-plane rotations offset by 60° (see SI, Fig. S3). While EBSD cannot inform on the relative orientation of the $MoS_2$ lattice to the Au lattice, we repeated the sample fabrication on a TEM grid to obtain the Fast Fourier transformation (FFT) diffractogram of a high-angle-annular-dark-field scanning transmission electron microscope (HAADF-STEM) image (Fig. 1h). The boundary between suspended $MoS_2$ and $MoS_2$/Au shows the relative crystallographic alignment of the two crystals. For this sample, the relative lattice registry of all $MoS_2$-Au boundaries at various pore edges was within ± 4°.

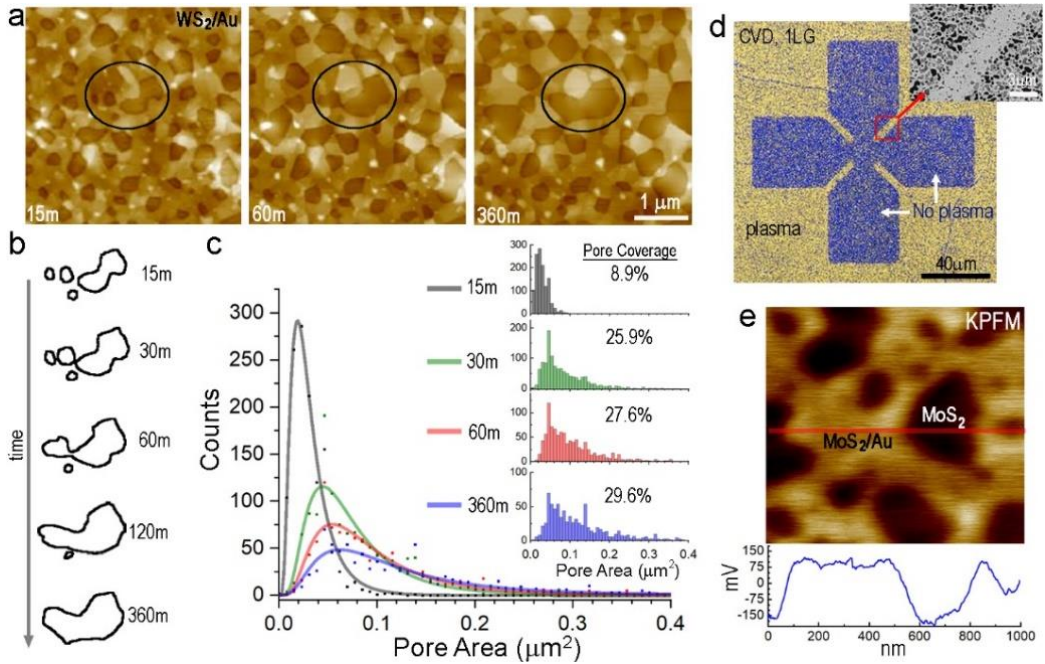

**Figure 2 | Evolution of 2DC/Au(25nm)/SiO$_2$ samples during annealing.** (**a**) Series of AFM images from the same WS$_2$/Au region after different cumulative annealing times (Ar+H$_2$, 300 °C; optical images of this region shown in Fig. 3d). (**b**) Outline of the pores inside the black ovals in **a** with increasing annealing time. (**c**) Plot showing the histogram peak value of the pore area *vs.* count for different cumulative annealing times (processed using ImageJ[26]). The solid lines are fits using the LSW particle-ripening model[27]. (inset) Full histograms of the pore area *vs.* count for different cumulative annealing times, along with the measured areal pore coverage for each annealing time. (**d**) Optical image showing a CVD graphene/Au/SiO$_2$ sample that has been selectively treated with a 15 sec O$_2$-plasma. After annealing, the Au layer beneath 'pristine' graphene dewets, while the Au beneath O$_2$-plasma exposed graphene has limited dewetting (inset scale bar = 3 μm) (**e**) KPFM image of a MoS$_2$/Au sample after annealing. The suspended MoS$_2$ membranes can have up to a 300 mV potential difference compared to the Au-supported MoS$_2$.

By closely monitoring Au dewetting beneath a 2DC layer, we gain insight into the dewetting dynamics. The evolution of metallic pore formation and growth under a WS$_2$ monolayer during cumulative annealing is illustrated in Figure 2a,b (optical image of the same region shown in Fig. 3d). The sequential annealing and imaging of the same location enables a quantitative analysis of changes in pore size, shape and frequency (or count) (Fig. 2c and SI, Fig. S4). We observe here that most of the pore formation and growth occurs within the first 60 min. of treatment. After an initial 15 min. anneal (300 °C) there is already a large density of small pores (<Area>= 0.028 μm$^2$, density= 3.13/μm$^2$), with a surface coverage of 8.9%. With further annealing, pores tend to grow, merge and exhibit well-defined facets, while the top Au surface is smooth with distinct step edges. At the longest anneal times (360 min) for this sample, the pore coverage increases to 29.6%, the average pore size increases by 5.5× (<Area>= 0.154 μm$^2$), and the pore density decreases by ~30% (density= 2.17/μm$^2$). Figure 2b highlights an example of these changes, where we outline individual pores circled in Fig. 2a as they change with cumulative annealing.

It is important to note the impact of defects in the 2DC layer on Au dewetting. Figure 2d shows an optical image of CVD graphene/Au film that has been selectively treated with oxygen plasma (15 sec) to intentionally introduce defects in the top graphene layer. When subsequently annealed, Au beneath pristine graphene dewets as expected ('bluish' region, Fig 2d), whereas Au dewetting beneath defective graphene is limited. We surmise that dangling bonds and functional groups in the graphene layer can effectively serve as pinning sites for Au diffusion, as well as change the local surface energy, and thereby impede the dewetting process. Using this result, we have selectively patterned the wetting/dewetting behavior of an underlying Au film with few-micron resolution (inset Fig. 2d and SI Fig. S5).

The growth of larger metal pores, seemingly at the expense of smaller ones (e.g., Fig. 2b), resembles a behavior analogous to Ostwald ripening[27] of solid particles. In this case, however, the particles under consideration are pores (i.e., absence of material). One of the more successful coarsening (or ripening) models– the so-called LSW theory named after Lifshitz and Slyozov,[28] as well as the work of Wagner[29]– uses a log-normal distribution to follow how particles lose or gain mass over time. When we apply a LSW model (Fig. 2c, solid lines) to the pore distributions, we find relatively good overlap ($R^2$ values ranging from 0.887 to 0.977) to the data, which suggests pore growth here follows conventional particle ripening kinetics. As mentioned earlier, the competing forces at play are the differences in surface energy between the metal and the substrate, and the metal and the 2DC layer, as well as the high Young's modulus of the 2DC, which reduces surface fluctuations that initiate dewetting.[24] Without the 2DC overlayer, the Au film completely dewets to form localized Au islands with randomly orientated in-plane crystal directions.

**Optical and electronic properties of 2DC/metal architectures.** Beyond morphological considerations, the OPEN film geometry leads to a rich mixture of electronic and optical properties across the 2DC layers. Specifically, a discrete distribution of nano- and micron-scale interface junctions exist after dewetting consisting of: (i) planar, crystallographically aligned 2DC/metal boundaries, (ii) sharp metallic pore edges and corners with a conformal 2DC coating, and (iii) strained and suspended 2DC layers. It is known that hybridized electronic states form at metal/2D semiconductor junctions, where the bandstructure of the semiconductor typically renormalizes.[30] We confirm such changes exist here using kelvin probe force microscopy (KPFM; Fig 2e) to measure surface potentials of the $MoS_2$/Au regions and free-standing $MoS_2$ regions. These two areas differ in surface potential by up to 300 mV, similar to other reports.[31] This variation primarily corresponds to band bending of $MoS_2$, due to the difference between the Au and $MoS_2$ work functions.[31,32] As a result, regions of suspended 2DC membranes in the OPEN film can serve as localized potential wells for charge carriers (e.g., electrons or holes depending on the metal/2DC junction) and exhibit a notable impact on emission properties, as described below.

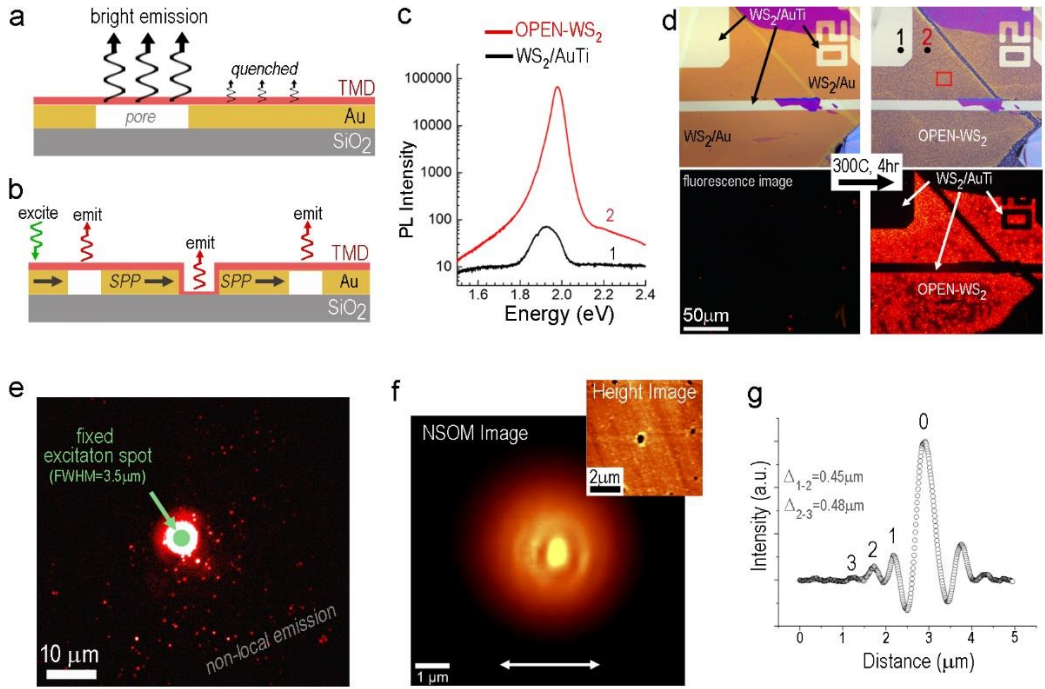

**Figure 3 | Optical characterization of OPEN-WS$_2$ samples (WS$_2$/Au(25nm)/SiO$_2$).** Cartoon illustrating (**a**) strong emission from the suspended TMD and quenched emission where it is supported by Au and (**b**) non-local excitation and emission from the an OPEN-WS$_2$ film on Au film. (**c**) PL spectra (log$_{10}$ intensity) taken from spots "1" and "2" in the optical image in **d**. (**d**) Optical and fluorescence images of a WS$_2$/Au/SiO$_2$ sample before and after annealing (red box = AFM region from Fig. 2a). Regions of Au-Ti were selectively patterned to inhibit dewetting at specific locations. (**e**) Fluorescence image of an OPEN-WS$_2$ film on Au with a fixed excitation laser spot (beam FWHM ≈ 3.5 μm, power= 4.5 mW, room temperature) demonstrating non-local excitation of suspended WS$_2$ pixels. (**f**) Topography and near-field interference pattern originating from SPPs (532 nm source focused from below the pore, power= 0.23 mW). (**g**) Horizontal cross-section of near-field intensity taken from **f**.

Variations in the optical properties (both local and non-local) across the OPEN film are equally striking and are the subject of the remaining analysis and discussions. Since light emission in 2D semiconductors is quenched on metal surfaces, we anticipate large intensity differences between the Au-supported and free-standing 2DC regions (Fig. 3a). Photoluminescence (PL) on an OPEN-WS$_2$ film and a supported WS$_2$/Au-Ti film shows that the free-standing OPEN-WS$_2$ membranes can have 1,000× brighter PL than the Au-supported regions (Fig. 3c). To visualize this on a larger scale, Figure 3d shows optical and fluorescence images of a WS$_2$/Au sample before and after annealing. Local regions of Au/Ti were patterned here as 'anchoring' sites (dewetting does not occur with a Ti sticking layer) to illustrate one avenue for directed dewetting. Immediately upon formation of an OPEN-WS$_2$ film, we observe fluorescence that intensifies with increased annealing time as more free-standing WS$_2$ is generated (Fig. 3d lower right).

Non-local excitation of PL, facilitated by SPP traveling waves in the metal framework, can also occur over tens of microns in these samples. Excitation transfer over these distances is shown in Fig. 3e, where bright emitters appear at pore sites up to about 20 μm from the illumination spot (3.5 μm FWHM). We propose excitation transfer via SPPs is responsible for this long-range excitation (schematically shown in Fig. 3b), similar to that observed in TMD/dielectric/metal geometries.[13] Generation and propagation of SPPs was imaged by near-field scanning optical microscopy (NSOM) and is shown in Fig. 3f. SPPs generated from free space excitation through a perforated metal film generate standing wave interference patterns whose periodicity match the SPP's in-plane wavelength.[33,34] Considering the dielectric function of Au, $\varepsilon_m$,[35] and incident excitation $\lambda_i$=532 nm, the SPP's wavelength would be $\lambda_{SPP} = \lambda_i\sqrt{(Re(\varepsilon_m) + 1)/Re(\varepsilon_m)} = 471\ nm$. This matches well with oscillatory fringes seen in a section profile taken through Fig. 3f and plotted in Fig. 3g. As expected for fringes associated with SPPs emanating from a nanopore in Au,[33] fringes are more pronounced along the direction of incident polarization (polarization direction is identified by the horizontal white arrow in Fig. 3f) and have further verified this property by rotating incident polarization (see SI). The propagation length (γ) of an SPP in our system, dominated by resistive losses in the metal, is estimated at approximately γ≈ 1 μm when excited by a 532 nm laser field.[33,36] This direct NSOM imagining clearly identifies the presence of SPPs launched from the myriad holes present in the metal framework and provides the basis for non-local excitation of bright emission centers.

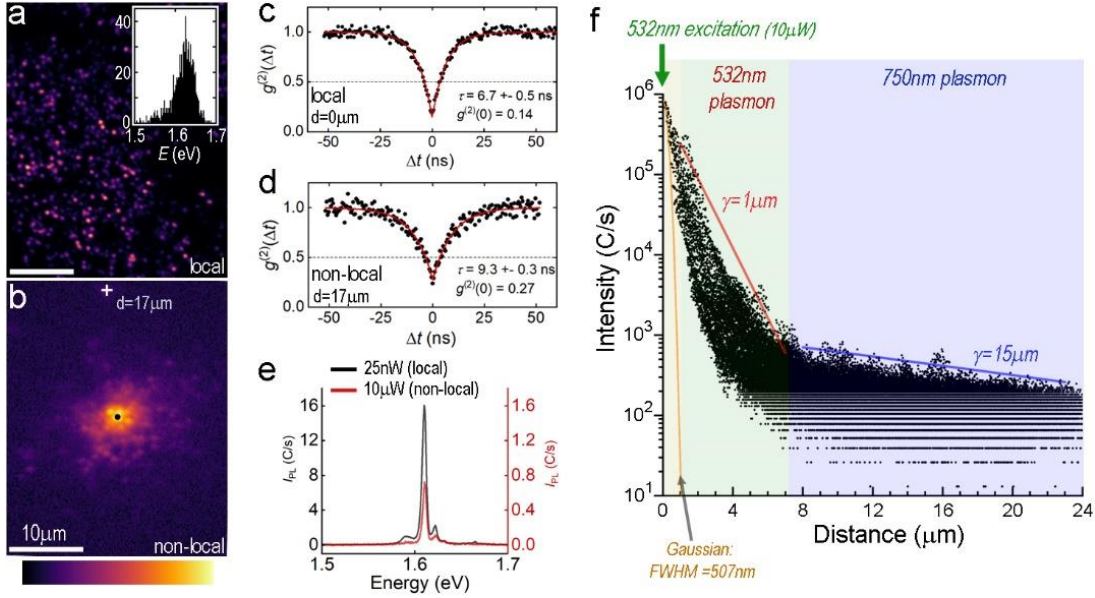

**Figure 4 | Low-temperature (4.5 K) characterization of an OPEN-WSe$_2$ film (Au/WSe$_2$/Au/SiO$_2$).**
(**a**) Scanning confocal PL map (linear color scale, 0 to 0.8×10$^5$ c/s) and (**b**) 'non-local' PL map (logarithmic color scale, 50 to 1.1×10$^5$ c/s) with the collection spot fixed at '●' point and the excitation beam scanned to form the image. (**c,d**) second-order photon correlation function $g^{(2)}(\Delta t)$ measurement of the central emitter in **b** for excitation at the '●' point and non-locally at '+' point, respectively. (**e**) PL spectra taken at the central emitter while exciting locally (excitation and collection at '●') and non-locally (excitation at '+', collection at '●'). The scale of the non-local spectrum (red curve) is on the right-hand y-axis. (**f**) Plot of PL peak intensity versus distance for every pixel in **b**. Several guide lines are added to the plot including a Gaussian curve representing the diffraction limited intensity of the laser spot (orange curve, 507 nm FWHM), and two exponential decay curves using a decay length of $\gamma = 1$ μm (red line) and $\gamma = 15$ μm (blue line). The green and blue colored regions highlight the excitation wavelength required to launch SPPs with the specified decay length.[36]

**Local and non-local quantum emission.** Photon-mediated interactions between discrete on-chip quantum light sources is the foundation of integrated quantum optics. We demonstrate key requirements toward this goal, including the remote excitation of single-photon emitters (SPEs). Using an OPEN-WSe$_2$ film, we measure a distribution of narrow emission centers and extract relevant statistics. Fig. 4a shows an example of a low-temperature scanning confocal PL map where the excitation and collection spots are coincident. A histogram of emitter peak energy from this map has a peak energy centered at 1.625 eV (inset Fig 4a). From this same region, we can collect 'non-local' PL maps to quantify aspects of remote excitation. In Fig. 4b we fix the collection spot at one emitter and scan the excitation spot to learn where energy is 'in-coupled', transmitted through the porous metal framework, and 'out-coupled' via the emitter under observation. The bright spots in Fig 4b. away from the center indicate the position of the excitation beam that results in sufficient energy in-coupling and transfer to result in significant emission central emitter.

Using this experimental geometry, we perform photon counting statistics to compare emitter response when excited by a direct laser field versus a propagating SPP, as well as quantify the distance-dependence of the remote PL. Figure 4c,d shows the measured second-order correlation function $g^{(2)}(\Delta t)$, where we find that locally $g^{(2)}_{d=0\mu m}(\Delta t = 0) = 0.14$ and non-locally at a distance of 17 μm $g^{(2)}_{d=17\mu m}(\Delta t = 0) = 0.27$. The individual emission spectra (Fig. 4e) for local (laser field) or non-local (SPP) excitation reveals no significant differences in overall spectral shape. Differences in the emitter radiative lifetime, as estimated from the width of the $g^{(2)}(\Delta t)$ antibunching dip, may be due to differences in the effective excitation power at the emitter site.

Finally, by plotting the intensity of every pixel versus distance from the central emitter in Fig. 4b, we can estimate the decay length of remote PL. Figure 4f shows a semi-log plot of the non-local PL intensity, together with two different decay length lines ($I \propto I_o e^{-x/\gamma}$) to guide the eye. Two distinct regions are evident in the PL intensity plot. Less than about 7 μm from the emitter, the PL intensity decays rapidly and aligns reasonably well with a decay length of $\gamma$ = 1 μm. Beyond 7 μm from the emitter, the decay rate significantly slows and aligns well with $\gamma$ = 15 μm. As discussed earlier, the estimated decay length of plasmons in Au excited by 532 nm light is $\gamma \approx$ 1 μm, which agrees well with the decay length within 7μm of the pump laser. A decay length of 15 μm in gold films with nanometer roughness would require an excitation energy of about 750nm,[36] which is precisely the emission energy of monolayer $WSe_2$. This result provides strong evidence that emitted light from $WSe_2$ can launch SPPs into the metal framework, which propagate over much longer distances across the gold film to excite $WSe_2$ quantum emitters remotely.

**CONCLUSION**

In summary, we have reported on the physical and electronic interplay between two-dimensional crystal (2DC) layers and self-assembled, highly-textured porous metallic frameworks. Annealing a 2DC/Au/$SiO_2$ stack results in a 'reverse epitaxy' process of the encapsulated Au layer, where the Au layer crystallographically orders with the 2DC overlayer and transforms into an orientated pore enabled network (OPEN) film. The 2DC overlayer remains suspended above or coats the inside of the metal pores, which serve as local windows to access intrinsic properties of the 2DC layer. When using a semiconducting 2DC layer, bright emission occurs at the 2DC/metal-pore sites and is quenched at the 2DC/metal regions. The porous metal framework supports propagating surface plasmon-polaritons (SPPs) launched by either a direct laser field or light emitted from the TMD semiconductor. These SPPs travel tens of microns in the metallic framework to re-excite excitons in a TMD overlayer. Using this process, we measure both local and non-local single photon emission from the same emitter site, which shows the quantum nature of the emission is preserved whether the excitation originates from a laser field or a propagating SPP.

## METHODS

**Sample fabrication.** To generate exfoliated 2DC/Au samples, Au films were first evaporated (Pressure ≤ 8E-7 torr; rate = 1Ang/s, room temperature) onto PVA coated Si substrates (PVA is used here as a sacrificial layer). Bulk crystals (graphene, $h$-BN, $MoS_2$, $WS_2$, and $WSe_2$) were then exfoliated onto the Au/PVA/Si substrates to produce large-area monolayer regions. A macroscopic scratch was drawn through the Au/PVA layer around regions of interest and then local water drops were added to dissolve and undercut the PVA layer. The released 2DC/Au films were re-deposited on $SiO_2$/Si substrates. Large-area graphene films were grown by low-pressure CVD using Cu foils at 1030°C with flowing $H_2$/$CH_4$ gas (total pressure between 20-100 mtorr) and subsequently transferred onto Au/$SiO_2$ substrates using wet chemical etching and a PMMA support coating. 2DC/Au film samples were subsequently annealed in a 1-inch 'clamshell' tube furnace with flowing Ar (600 sccm) and $H_2$ (400 sccm) gas between 15-360 minutes, then quenched by opening the furnace.

**Near-field scanning optical microscopy (NSOM).** NSOM measurements were carried out on a Witec alpha300 in collection mode. Light (532 nm laser) is incident through the bottom of the sample. The illumination optic is fixed to the scanning stage so that the illuminated area does not change during scanning. Light is collected through an apertured AFM tip (~90 nm aperture size, contact mode) and measured with an avalanche photodetector (APD).

**Variable-temperature confocal photoluminescence.** Variable temperature PL measurements (4.5 K - RT) were acquired in a closed cycle optical cryostat with an internally mounted 100×, 0.85 NA microscope objective. A 532 nm diode-pumped solid state laser was used for excitation. Emitted light from the sample was passed through a 600 nm long pass filter and then coupled into an optical fiber for routed either to a spectrometer or to a beamsplitter and two APDs for photon correlation measurements. Scanning confocal PL images and split excitation-collection PL images were collected using two dual-axis scanning mirrors and two relay lens pairs, allowing separate control of both the laser excitation and confocal collection points on the sample.

**Electron backscatter diffraction (EBSD).** EBSD was carried out in a Helios Focused Ion Beam system. For the averaged diffraction plots in Figure 1g, data was collected from $60\times40\mu m^2$ areas on each region (same sample). Diffraction patterns were acquired approximately every 1.3μm totaling approximately 1,200 data points.

**Scanning transmission electron microscopy.** High-angle-annular-dark-field (HAADF) scanning transmission-electron-microscope (STEM) images were acquired with an aberration-corrected STEM (Nion UltraSTEM200X) operated at 60 kV with a probe current of about 50 pA. The STEM samples were prepared by transferring 2DC/Au films to commercial heating substrates coated with thin carbon films that contain arrays of 2 µm holes (Protochips E-FHBC). The as-transferred 2DC/Au films were then directly annealed on these substrates to form an 'OPEN' film before insertion into microscope for imaging.

## ACKNOWLEDGEMENTS

This work was funded by the Office of Naval Research through Base Programs at the Naval Research Laboratory. MOCVD materials used in this study were provided by The Pennsylvania State University 2D Crystal Consortium − Materials Innovation Platform (2DCC-MIP) under NSF cooperative agreement DMR-1539916.